# Accuracy and Efficiency of Simplified Tensor Network Codes

David Yevick and Jesse Thompson

Department of Physics

University of Waterloo

Waterloo, ON N2L 3G7

**Abstract:** We examine in detail the accuracy, efficiency and implementation issues that arise when a *simplified* code structure is employed to evaluate the partition function of the two-dimensional square Ising model on periodic lattices though repeated tensor contractions.

**Introduction:** Because the number of configurations grows rapidly with system size, the exact determination of thermodynamic (macroscopic) quantities such as the free energy, magnetization or specific heat of a lattice model requires prohibitively large computational resources for all but the smallest system sizes. Accordingly, both stochastic and deterministic approximation procedures have been developed to enable the analysis of physically interesting systems. Stochastic methods, which employ probabilistic techniques to sample the configuration space, include the Monte-Carlo, Metropolis, [1] multicanonical, [2] [3] [4] Wang-Landau [5] [6] [7] and transition matrix [8] [9] [10] [11] [12] [13] procedures. While the last of these approaches is perhaps the most flexible and accurate technique, In the presence of phase transitions, however, its efficiency is degraded by correlations between successive samples. While this difficulty can be partly surmounted by monitoring the sampling of appropriate configuration space regions [12] or by applying cluster reversal algorithms, [14] a large number of samples are still required to achieve suitable accuracy for quantities such as the specific heat that involve multiple derivatives of the partition function..

Tensor network algorithms, [15] [16] [17] [18] [19] [20] [21] [22] [23] in contrast, constitute an efficient and relatively straightforward strategy for coarse-graining lattice problems in an approximate yet controllable manner. The technique expresses the partition function at a given temperature as a product of tensors of which the $i$:th term, $T^{(i)}$, incorporates all subsystem configurations at the $i$:th lattice site. The contribution to the partition function associated with the interaction between two adjacent sites is evaluated by contracting (multiplying) their common index or indices. While the number of products that would need to be summed over (contracted) in this manner increases exponentially with the number of sites, the computational cost can be limited by performing a singular value decomposition of the matrix formed by the elements that enter into contraction and retaining only the subspace spanned by the basis vectors associated with the largest singular values. However, in contrast to stochastic methods, which trivially extend to grids of any geometrical structure, tensor procedures cannot be simply adapted to heterogeneous networks.

Tensor network algorithms are often analyzed in the thermodynamic limit of periodic systems for which the effect of the system boundaries becomes negligible. The magnitude of the tensor components then attain stable, boundary independent values since each contraction step effectively defines a renormalization transformation of the lattice. However, in many practical applications the system size is relatively small so that boundary effects are significant. While several authors have recently examined the accuracy of sophisticated tensor network algorithms for arbitrary system sizes, [16] [18] [24] [25] [26], a presentation of the method based on a compact programming strategy that is accessible to non-specialists does not, to our knowledge, exist at present.

Accordingly, in this paper we extend the analysis of [27] by applying a simplified tensor network procedure to the spin ½ square lattice Ising model benchmark calculation. We consider systems with up to $64 \times 64$ lattice sites for which we possess exact solutions for comparison. We find that a high degree of precision can be achieved at all temperatures if enough singular values are retained. However, the computation time required to reach a fixed level of accuracy increases rapidly with the number of spins. Our presentation complements previous studies of the two-dimensional Ising model in that we present and explain all essential computer code and further fully analyze the convergence and efficiency of specific heat calculations.

**Computational Methods:** The analysis of the square Ising model is best illustrated by presenting the central equations of the tensor network method in the context of corresponding program excerpts and pictorial diagrams. To apply the procedure, a lattice Hamiltonian must be first expressed, generally though a transformation to the dual lattice, in terms of the degrees of freedom on the lattice bonds. This step must be formulated in such a manner that the Boltzmann weight of a configuration transforms into a product of Boltzmann weights at individual sites, each of which depends solely on the adjoining bond variables. In the present context, recasting the two-dimensional Ising model on a square lattice of $N \times N$ spins with periodic boundary conditions in terms of bond variables in the dual lattice yields the modified partition function

$$\tilde{Z} = tTr\left[\otimes_{i=1}^{N^2} T^{(i)}\right] \tag{1}$$

where $T^{(i)}$ represents the local tensor at the $i$:th lattice site. The definition of the tensor trace operator, $tTr$, in this expression is however somewhat abstract since it implies contraction over each pair of indices that are connected through nearest neighbor interactions as a result of the lattice structure and boundary conditions, as will become apparent in the discussion below. The physical partition function is related to the modified partition function through

$$Z = \tilde{Z}(\sinh 2K_p)^{-N^2} \tag{2}$$

while the tensor $T^{(i)}$ is given for the two dimensional square Ising model by

$$T^{(i)}_{lurd} = \frac{1 + p_l^{(i)} p_u^{(i)} p_r^{(i)} p_d^{(i)}}{2} e^{K_p\left(p_l^{(i)} + p_u^{(i)} + p_r^{(i)} + p_d^{(i)}\right)/2} \tag{3}$$

The quotient term in Eq.(3) corresponds to a projection operator that equals unity for the allowed bond configurations and is zero otherwise. To implement the tensor formalism we will employ the python **tensornetwork** package [28] [29] [30] imported as **net,** for which the leftward bond, $l$, emanating from node $p$ is assigned the index 0, while the upper, right and downward bonds, $u, r, d$ are given indices 1,2,3, respectively, as indicated in Figure 1. Each of the bond variables, $p_\alpha^{(i)}$, can equal either 1 or 0, corresponding to anti-aligned and aligned spins on either side of the bond, that respectively contribute $-J < 0$ and $J$ to the total system Hamiltonian where $J$ is the Ising model spin-spin coupling strength. The dual lattice variable $K_p$ is expressed in terms of the dimensionless quantity $K = \beta J = J/k_B T$, by

$$K_p = \frac{\log(\tanh K)}{2} \tag{4}$$

We employ units for which $J/k_B = 1$, effectively measuring the temperature in units of $J/k_B$. The code that constructs these tensors and then assigns them to nodes of a square network, following the naming and formatting standards of [31] [32] is (after importing **product** from **itertools**):

```
KP = -0.5 * np.log( np.tanh ( K ) )
myNode = [ [ None ] * numberOfRows for i in range( numberOfRows ) ]
T = np.zeros([D, D, D, D])
for r, u, l, d in product( range( D ), repeat = 4 ):
   T[r][u][l][d] = 0.5 * ( 1 + (2 * l - 1) * (2 * u - 1) * (2 * r - 1) * (2 * d - 1) ) * \
   np.exp( KP * ( l + u + r + d - 2 ) )
for rowLoop in range( numberOfRows ):
    for columnLoop in range( numberOfRows ):
       myNode[rowLoop][columnLoop] = net.Node( T, f"node({rowLoop}, {columnLoop})" )
```

Once the node objects have been instantiated and stored in a two-dimensional list, they are linked by joining, for example, the rightward pointing, 2, bond of each lattice site to the leftward pointing, 0, bond of its right neighbor site. The boundary conditions at the edges of the lattice are implemented by identifying the right neighbor of the rightmost lattice site of a given row with the leftmost node of the row. A similar procedure is applied to the vertical bonds. The code for this operation is:

```
for rowLoop in range( numberOfRows ):
   for columnLoop in range( numberOfRows ):
      rowLoopP1 = ( rowLoop + 1 ) % numberOfRows
      columnLoopP1 = ( columnLoop + 1 ) % numberOfRows
      net.connect( myNode[rowLoop][columnLoop][3], myNode[rowLoopP1][columnLoop][1] )
      net.connect( myNode[rowLoop][columnLoop][2], myNode[rowLoop][columnLoopP1][0] )
```

This results in the structure depicted in the tensor network diagram of Figure 1, where each dot and the line between each pair of dots correspond to a node tensor and to the shared index between the two node tensors, respectively.

While the number of spin configurations that must be summed over to generate the partition function grows exponentially with the number of nodes of the lattice, singular value decomposition can be employed to restrict the summations to the combinations of $D_{cut}$ tensor elements that yield the greatest contribution to the partition function. For a regular lattice this enables an iterative procedure in which a new lattice is obtained with the same geometry but a smaller number of nodes. After reducing the original lattice to a single node an approximation to the partition function is obtained with an accuracy principally determined by the value of $D_{cut}$. However, the singular value decompositions increase the number of elements of the node tensors from $2^4$ to $D_{cut}^4$. Consequently, the required computational resources increase markedly with $D_{cut}$.

The reduction from a square lattice with $N^2$ nodes, where $N$ is a power of 2, to a second lattice with $(N/2)^2$ nodes proceeds in two steps as illustrated in Figure 2 - Figure 4. First, the SVD is applied to recast the tensors on alternate lattice sites as products of the form, where the self-evident superscripts $(2i)$ or $(2i+1)$ of the $S$ tensors are omitted,

$$T^{(2i)}_{lurd} = \sum_{\alpha=1}^{D_{cut}} \hat{S}^0_{lu,\alpha} \hat{S}^1_{rd,\alpha}$$
$$T^{(2i+1)}_{lurd} = \sum_{\alpha=1}^{D_{cut}} \hat{S}^0_{ur,\alpha} \hat{S}^1_{dl,\alpha} \tag{5}$$

The associated program lines are:

```
for rowLoop in range( numberOfRows ):
   for columnLoop in range( numberOfRows )
      rowLoopM1 = ( rowLoop - 1 ) % numberOfRows
      rowLoopP1 = ( rowLoop + 1 ) % numberOfRows
      columnLoopP1 = ( columnLoop + 1 ) % numberOfRows

      if ( rowLoop + columnLoop ) % 2 == 0:

         splitNode[rowLoop][columnLoop][0], splitNode[rowLoop][columnLoop][1], error = \
         net.split_node( myNode[rowLoop][columnLoop], \
         [myNode[rowLoop][columnLoop][0], myNode[rowLoop][columnLoop][1]], \
         [myNode[rowLoop][columnLoop][2], myNode[rowLoop][columnLoop][3]], \
         max_singular_values = myMaximumSingularValues )

      else:

         splitNode[rowLoop][columnLoop][0], splitNode[rowLoop][columnLoop][1], error = \
         net.split_node( myNode[rowLoop][columnLoop], \
         [myNode[rowLoop][columnLoop][1], myNode[rowLoop][columnLoop][2]], \
         [myNode[rowLoop][columnLoop][3], myNode[rowLoop][columnLoop][0]], \
         max_singular_values = myMaximumSingularValues )
```

Mathematically, the four component tensor $T^{(i)}_{lurd}$ is first expressed as either the $D^2 \times D^2$ matrix $T^{(2i)}_{\{lu\}\{rd\}}$ or as $T^{(2i+1)}_{\{ul\}\{dl\}}$, where initially $D = 2$. The matrix $S$ that yields the best-fit approximation in Eq.(5) is then obtained by minimizing the norm $|T - SS^T|$. This is achieved through the singular valued decomposition, $T^{(i)}_{\{mn\},\{pq\}} = \sum_{i=1}^{D^2} s_i U_{\{mn\},i} V^*_{\{pq\},i}$ in which the $s_i$ are the singular values and $U$ and $V$ are unitary matrices. Subsequently only the only the columns $U$ and $V$ that correspond to the largest $D_{cut}$ singular values are retained such that, denoting truncation by a hat symbol, $\hat{S}^0_{i,\{mn\}} = \sqrt{s_i}\hat{U}_{\{mn\},i}$ and $\hat{S}^1_{i,\{pq\}} = \sqrt{s_i}\hat{V}^*_{\{pq\},i}$ in Eq.(5).

Subsequently, contracting over the squares in the tensor network diagram of Figure 2, generates a new tensor for each square according to

$$\hat{T}^{(i)}_{lurd} = \sum_{\alpha,\beta,\gamma,\delta=1}^{D_{cut}} \hat{S}^1_{\delta\alpha,l} \hat{S}^1_{\alpha\beta,u} \hat{S}^0_{\beta\gamma,r} \hat{S}^0_{\gamma\delta,d} \qquad (6)$$

The $N \times N$ square lattice is consequently transformed into a $N \times (N/2)$ diamond lattice which is stored as a two-dimensional rectangular list array. The **tensornetwork** code is given by,

```
for rowLoopO2 in range( int( numberOfRows / 2 ) ):
    for columnLoopO2 in range( int( numberOfRows / 2 ) ):

        rowLoop = 2 * rowLoopO2
        columnLoop = 2 * columnLoopO2
        columnLoopP1 = ( columnLoop + 1 ) % numberOfRows
        rowLoopP1 = ( rowLoop + 1 ) % numberOfRows
        columnLoopP2 = ( columnLoop + 2 ) % numberOfRows
        rowLoopP2 = ( rowLoop + 2 ) % numberOfRows

        newNode[rowLoop][columnLoopO2] = splitNode[rowLoop][columnLoop][1] @ \
        splitNode[rowLoop][columnLoopP1][1] @ splitNode[rowLoopP1][columnLoopP1][0] \
        @ splitNode[rowLoopP1][columnLoop][0]
        newNode[rowLoopP1][columnLoopO2] = splitNode[rowLoopP1][columnLoopP1][1] @ \
        splitNode[rowLoopP1][columnLoopP2][1] @ splitNode[rowLoopP2][columnLoopP2][0] \
        @ splitNode[rowLoopP2][columnLoopP1][0]
```

The modulus operator, **%**, again enforces the periodic boundary conditions that generate the last row of nodes in the rectangular lattice.

The reduction of the rectangular lattice to a $(N/2) \times (N/2)$ square lattice is performed analogously as illustrated by the tensor network diagrams shown in Figure 3 and the associated code segment, which follows the numbering convention of Figure 4.

```
for rowLoop in range( numberOfRows ):
    for columnLoop in range( int( numberOfRows / 2 ) ):

        if rowLoop % 2 == 0:
            splitNode[rowLoop][columnLoop][0], splitNode[rowLoop][columnLoop][1], error \
            splitNodeError[rowLoop][columnLoop] = \
            [newNode[rowLoop][columnLoop][0], newNode[rowLoop][columnLoop][3]], \
            [newNode[rowLoop][columnLoop][1], newNode[rowLoop][columnLoop][2]], \
            max_singular_values = myMaximumSingularValues )
        else:
            splitNode[rowLoop][columnLoop][0], splitNode[rowLoop][columnLoop][1], error \
            splitNodeError[rowLoop][columnLoop] = \
            [newNode[rowLoop][columnLoop][2], newNode[rowLoop][columnLoop][3]], \
            [newNode[rowLoop][columnLoop][0], newNode[rowLoop][columnLoop][1]], \
            max_singular_values = myMaximumSingularValues )

for rowLoopO2 in range( int( numberOfRows / 2 ) ):
    for columnLoop in range( int( numberOfRows / 2 ) ):

        rowLoop = 2 * rowLoopO2
        columnLoopP1 = ( columnLoop + 1 ) % int( numberOfRows / 2 )
```

```
      rowLoopP1 = ( rowLoop + 1 ) % numberOfRows
      rowLoopM1 = ( rowLoop - 1 ) % numberOfRows
      myNode[rowLoopO2][columnLoop] = splitNode[rowLoop][columnLoop][1] @ \
      splitNode[rowLoopM1][columnLoop][0] @ splitNode[rowLoop][columnLoopP1][0] @ \
      splitNode[rowLoopP1][columnLoop][1]
```

Since the new network is again square, iterating the above procedure $\log_2 N$ times produces a single node. However, for large $N$, the rapid growth in the tensor network elements associated with the exponentially large number of configurations induces numerical overflow unless the tensor elements are periodically rescaled. [33] This is conveniently implemented at the end of each iteration by

```
numberOfRows = int( numberOfRows / 2 )
if numberOfRows > 1:
    for rowLoop in range( numberOfRows ):
        for columnLoop in range( numberOfRows ):
            myNode[rowLoop][columnLoop].tensor = myNode[rowLoop][columnLoop].tensor / \
            myNode[0][0].tensor[0][0][0][0]
```

The physical partition function is finally evaluated and returned to the calling program through

```
Z = myNode[0][0]
result = Z.tensor * np.sinh( 2 * KP )**-numberOfSites
return result
```

Our evaluation of the specific heat per lattice site then employs the standard three-point finite difference discretization

$$c = \frac{1}{N^2}\frac{1}{kT^2}\frac{d^2 Z}{d\beta^2} \approx \frac{1}{N^2}\frac{1}{kT^2}\frac{Z(\beta + \Delta\beta) - 2Z(\beta) + Z(\beta - \Delta\beta)}{(\Delta\beta)^2} \quad (7)$$

Since numerical errors tend to increase with $N$, Eq.(7) was evaluated with several values of $\Delta\beta$ to ensure that the results for all $N$ considered are effectively independent of the value of this parameter.

**Results:** To benchmark the above procedure, we calculate the specific heat of $N \times N$ square Ising lattices with periodic boundary conditions for $N = 8, 16, 32, 64$. That the method is accurate over the entire inverse temperature region for $1.5 < \beta < 3.5$ is demonstrated in Figure 5, which compares the exact $N = 64$ specific heat obtained as in [34] [35] (solid line) to the tensor network calculation for $D_{cut} = 16$ with a finite difference point spacing of $\Delta\beta = 0.0025$ in Eq.(7) (+ markers). The maximum specific heat value generated in this computation is $c = 2.252047$ which agrees well with the result of the exact procedure, $c_{exact} = 2.252056$ Decreasing $\Delta\beta$ to 0.001 however yields large errors at a small number of equally spaced inverse temperatures for $1/T > 2.3$. Evidently, this numerical error, which becomes more pronounced as $D_{cut}$ is decreased, results from the interplay of the SVD truncation and the finite difference error. Large errors also result if $D_{cut}$ is substantially decreased. Thus, although excellent accuracy is obtained for properly chosen input parameters, the sensitivity of the calculation to these parameters can limit the practical application of our straightforward code to large systems.

Next, the dependence of the logarithm of the time required to evaluate the specific heat at the critical point for an infinite spin system, $1/T_c = 2.24$, on $N$ for $D_{cut} = 6$ is displayed as the dots in Figure 6. The solid line in the figure represents a linear fit given by $\log t = 1.33N - 5.35$. The power law exponent, 1.33, is significantly less than, for example, the value, $\approx 1.75$, reported for the optimal stochastic simulated annealing / transition matrix procedure of [11]. Together with Figure 5 this validates the applicability of the procedure to specific heat calculations with realistic boundary conditions in both the presence and absence of phase transitions. It should be noted, however that Figure 6 severely understates the growth with $N$ of the time required to reach a specified level of accuracy as $D_{cut}$ must then be simultaneously increased. This quantity, however, depends on the definition employed for the error and is additionally affected for larger values of $D_{cut}$ by finite difference and tensor contraction inaccuracies that are difficult to quantify individually. Accordingly, to isolate the contribution to the error associated with $D_{cut}$, Figure 7 displays the magnitude of the difference between the maximum value of the specific heat of each of the $N = 8, 16, 32, 64$ curves with $D_{cut} = 17$ and the corresponding values for each $N$ obtained with $2 \leq D_{cut} \leq 16$. While the considerable errors for $N = 32, 64$ are presumably largely suppressed in a quadruple precision calculation [33], this would clearly impact both the algorithmic performance and the programming simplicity.

**Discussion and Conclusions:** This paper has analyzed in detail the convergence of a simply coded tensor network procedure in the context of benchmark Ising model specific heat calculations with periodic boundary conditions. In future work, it would be of considerable interest to examine as well the programming effort, computational time and accuracy enhancements associated with the many available refinements to the method as the finite boundary effects could affect their performance at relatively small system sizes.

While the accuracy and efficiency of the tensor network procedure clearly exceeds those of stochastic methods for uniform lattices, it should be noted that the essential simplifying feature of the method, namely the iterative reduction of the grid size, is absent in heterogeneous lattices. That is, in the latter case, after contracting adjacent tensors a new lattice structure is obtained resulting in a considerable increase in program complexity. In contrast, statistical procedures can be applied to any lattice geometry without modification and are therefore generally preferable in these contexts.

**Acknowledgements:** The authors especially thank Caleb Cook for sharing with us the program described in [27] that initiated our examination of this topic. The Natural Sciences and Engineering Research Council of Canada (NSERC) is further acknowledged for financial support.

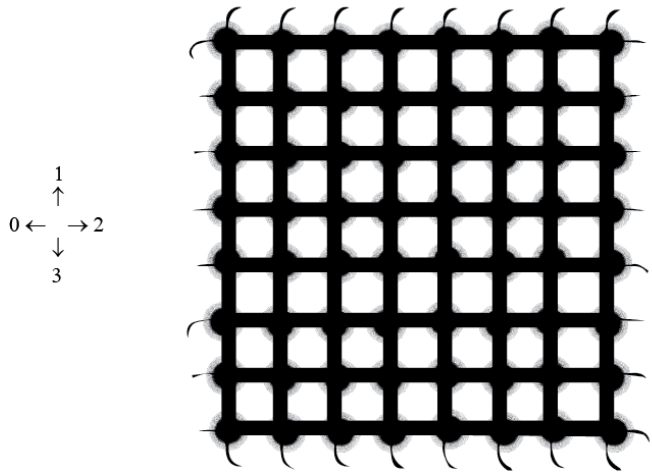

*Figure 1: A tensor network diagram for the partition function of the finite size Ising model with periodic boundary conditions. The circles at each of the vertices represent four index tensors while the lines indicate tensor contractions. The numbering convention for the bonds of a node employed in the python **tensornetwork** package is shown at the left of the figure.*

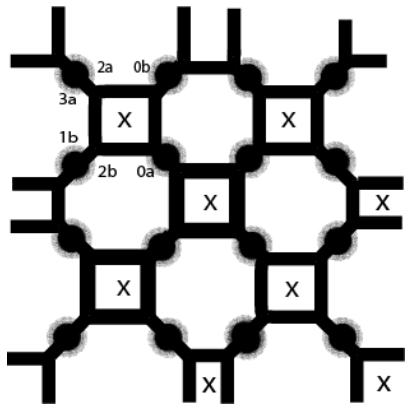

*Figure 2: The first step in contracting a 4 × 4 Ising model with periodic boundaries. The circles represent the original lattice sites while the new sites of the 4 × 2 lattice after the squares in the diagram are contracted are denoted by ×.*

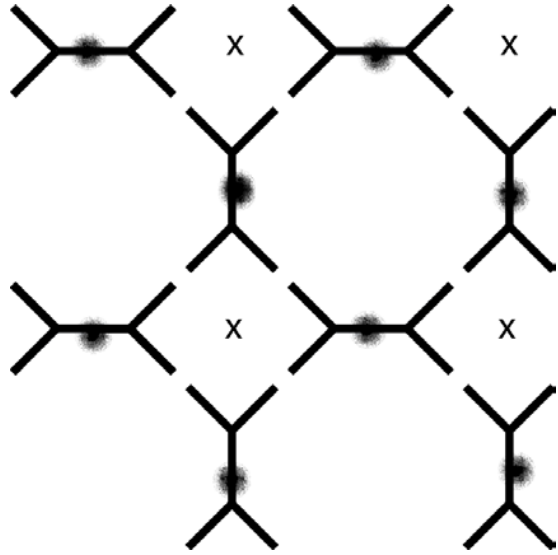

Figure 3: The second step in contracting the 4 × 2 lattice represented by circles into a new 2 × 2 square lattice (×) that serves as the starting point of a new iteration. The contractions are again performed over the squares in the diagram (see also following figure).

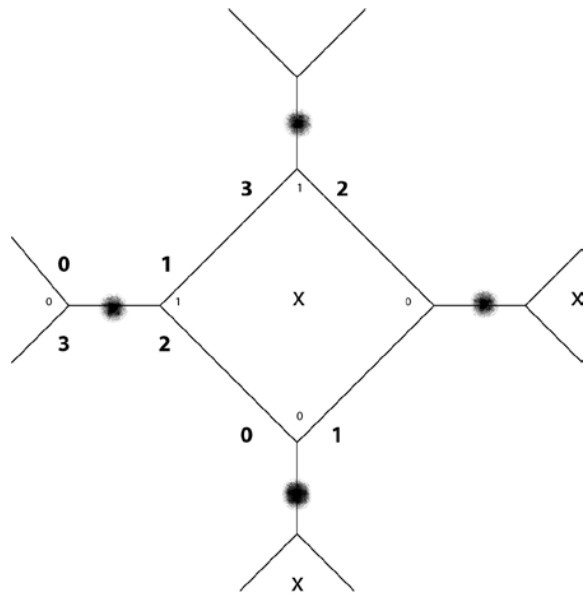

Figure 4: The numbering conventions employed in the previous diagram

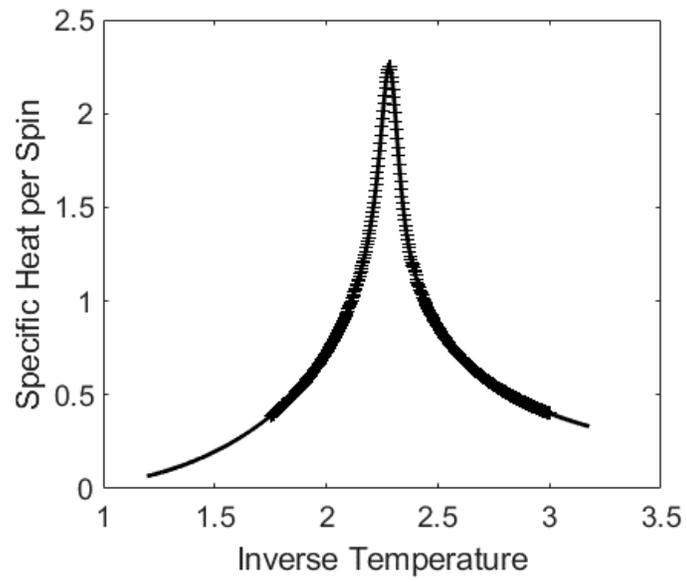

*Figure 5: The exact specific heat for the 64 × 64 Ising model (solid ine) together with the result of a tensor network calculation (+ markers)*

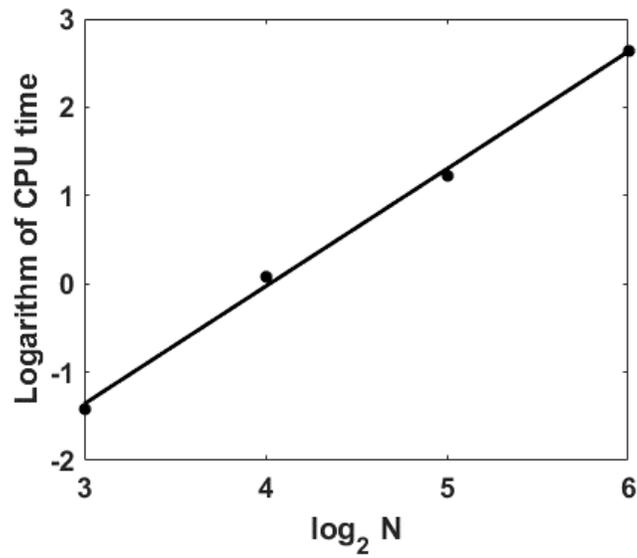

*Figure 6: The logarithm of the CPU time required for a single tensor network calculation of the specific heat for the N × N Ising model at the infinite lattice critical point for $D_{cut} = 6$.*

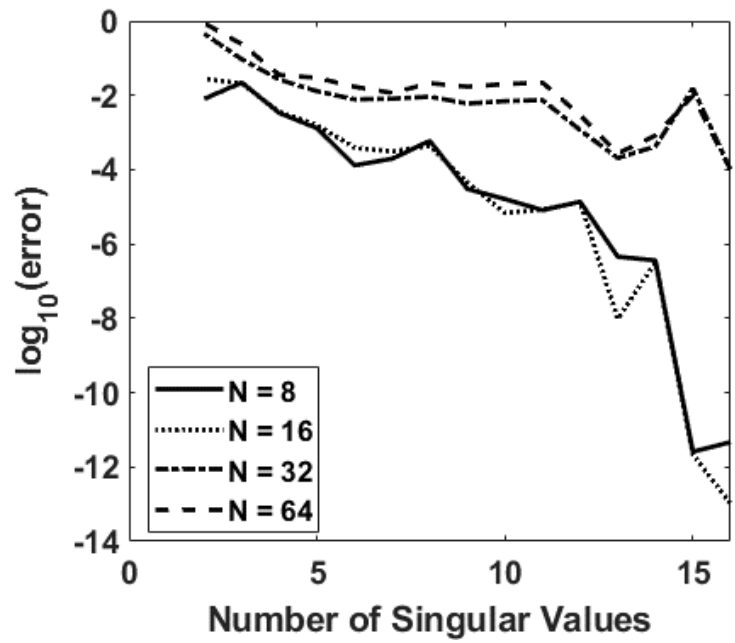

*Figure 7: The logarithm of the difference between the tensor network specific heat at the maximum of the specific heat curve and the value obtained for $D_{cut} = 17$ for $N = 8, 16, 32, 64$ as a function of $D_{cut}$.*